\documentstyle[floats,preprint,aps,epsf]{revtex}
\tightenlines

\newcommand{\hs}[1]{\hspace*{#1}}
\newcommand{\del}{\partial}
\newcommand{\nn}{\nonumber}
\newcommand{\cov}{\bigtriangledown}

\newcommand{\beq}{\begin{equation}}
\newcommand{\eeq}{\end{equation}}
\newcommand{\beqa}{\begin{eqnarray}}
\newcommand{\eeqa}{\end{eqnarray}}
\newcommand{\sr}{\sqrt}
\newcommand{\fr}{\frac}
\newcommand{\mn}{\mu \nu}

\begin{document}

\draft
\preprint{ KEK-TH-762, hep-th/0104213}

\title{Stable Black Strings in Anti-de Sitter Space}
\author{ Takayuki Hirayama\footnote{E-mail address: 
thirayam@post.kek.jp} and Gungwon Kang\footnote{E-mail address:
gwkang@post.kek.jp}} 
\address{ Theory Group, KEK, Tsukuba, Ibaraki 305-0801, Japan}
\maketitle
\begin{abstract}

In the five-dimensional Einstein gravity with negative cosmological 
constant in the presence/absence of a {\it non-fine-tuned} 3-brane, 
we have investigated the classical stability of black string 
solutions which are foliations of four-dimensional 
$AdS$/$dS$-Schwarzschild black holes. Such black strings are generically
unstable as in the well-known Gregory-Laflamme instability. For $AdS$ 
black strings, however, it turns out that they become stable 
if the longitudinal size of horizon is larger than the order of the 
$AdS_4$ radius. Even in the case of unstable black strings, the $AdS$ 
black strings have a very different feature of string fragmentations 
from that in the flat brane world. Some implications of our results on
the Gubser-Mitra conjecture are also discussed.

\end{abstract}


\bigskip

\newpage

\section{Introduction}

The Schwarzschild black hole which has a compact black hole horizon is 
known to be stable in general relativity~\cite{VRW}. However, it has 
been known that black string/brane solutions which are a foliation of 
lower dimensional Schwarzschild black holes are unstable, the so-called 
Gregory-Laflamme instability~\cite{GL}. This instability extends to a 
much broader range of charged black branes in string theory with the 
exception for extremal or near extremal cases~\cite{GLcharge,Reall}. 

In the presence of a cosmological constant, Gregory~\cite{Gcosmo} has 
for the first time shown that this black string instability persists 
to hold. Recently, Randall and Sundrum~\cite{RS} have proposed an 
interesting model where the gravity in higher dimensions with negative 
cosmological constant is localised on a lower dimensional domain wall. 
In the flat brane world model where the tension of a 3-brane is 
fine-tuned with the five-dimensional cosmological constant, any 
Ricci-flat four-dimensional metric can be embedded.
For instance, a black string solution, which is simply a foliation of 
four-dimensional Schwarzschild black holes perpendicular to the 3-brane,
can be easily constructed. The properties of such black string were 
investigated in Refs.~\cite{CHR,EHM}. In particular, they argued that 
the black string is unstable near the $AdS_5$ horizon, but becomes 
stable in the vicinity of the 3-brane, indicating the fragmentation 
into a cigar type (or a pancake-like more accurately~\cite{EHM}) black 
hole across the brane as a final state. Gregory~\cite{Gcosmo} has 
confirmed this conjecture by explicitly performing the linearized 
perturbation analysis of the black string background in this flat brane 
world scenario. The author also showed the instability of a black string 
embedded in the $AdS_5$ spacetime without the 3-brane. 

As long as we know, however, all black string/brane instabilities 
mentioned above were shown within the context of asymptotically locally 
flat spacetimes in the sense that all slices orthogonal to the string or
brane are asymptotically flat.\footnote{Actually Emparan, Horowitz, and
Myers have argued, based on entropy comparison, the stability of `BTZ'
black string solutions in {\it four-dimensional} $AdS$
space~\cite{EHM2}. However, they put two 2-branes with a suitable
identification of spacetimes. It seems that, even if the separation of
two 2-branes is taken to be infinitly large, this black string does not
correspond to a one dimension less configuration of our $AdS$ case with
a single 3-brane due to the specific embedding of the 2-branes.} 
Actually, in the presence of a negative
cosmological constant, it is also possible to construct black string
solutions which are a foliation of black holes that are asymptotically
{\it non-flat}. For example, the pure $AdS_5$ spacetime can be sliced 
into pure four-dimensional anti-de Sitter ($AdS_4$) or de Sitter
($dS_4$) spacetimes. Then, it is straightforward to show that any
four-dimensional metric satisfying four-dimensional Einstein equations
with the same cosmological constant can be embedded into $AdS_5$
space. In particular, the $AdS_4$ ($dS_4$)-Schwarzschild black hole can
be embedded, resulting in a five-dimensional hypercylindrical $AdS$
($dS$) black string solution. Similarly, even in the curved brane world
models where a 3-brane with non-fine-tuned tension is introduced in
$AdS_5$ backgrounds~\cite{KR,CBW}, this generalisation holds (see
Ref.~\cite{GS} for the $dS$ embedding.). 

Now it will be of interest to see whether the Gregory-Laflamme
instability still persists to hold for these black string solutions in 
$AdS$ backgrounds. It is because some naive arguments given below
seem to indicate that the stability behavior for such black strings
could be very different from the known Gregory-Laflamme instability. 
In addition, recently Gubser and Mitra~\cite{GM} have proposed an 
interesting conjecture about the relationship between the classical 
black string/brane instability and the local thermodynamic
stability~\cite{Reall}. It states that a black string/brane with a
non-compact translational symmetry is classically stable if, and only
if, it is locally thermodynamically stable. Since black string solutions
in $AdS$ space have warped geometries along the extra dimension, our 
study will show what will happen when the assumption of translational 
symmetry is discarded in the Gubser-Mitra (GM) conjecture. 

In this paper, we investigate the stability of black string solutions
which are asymptotically locally $AdS_4$/$dS_4$ mentioned above. 
It turns out that such black strings are generically unstable as usual.
Interestingly $AdS$ black strings, however, become stable if the
four-dimensional horizon radius is larger than the order of $AdS_4$
radius. This stability can be understood naively since the geometry along 
the string produces a sort of an effective compactification whose scale 
is determined by the $AdS_4$ radius. Even in the case of unstable $AdS$
black strings, the feature of instability is very different near the
conformal infinity. In fact, they become stable near the boundary of 
$AdS_5$. We will first present some naive arguments indicating this 
behavior by using entropy comparisons and the Gubser-Mitra conjecture. 
Then, linearized metric perturbation analysis and numerical results are 
shown explicitly. Finally, some discussions and physical implications of 
our results are followed.

\section{Black string solutions and naive stability arguments}
\label{naive}

Let us consider the five-dimensional pure anti-de Sitter spacetime
whose metric is given in the following form~\cite{KR,CBW} 
\beq
ds^2 = H^{-2}(z) (\gamma_{\mn}dx^{\mu}dx^{\nu} + dz^2 ).
\label{metric0}
\eeq
Here the warping factors are
\beqa
\label{warp1}
dS_4 (\Lambda_4 > 0) & : &  \qquad
H(z)= l_4/l_5 \sinh z/l_4  \\
\label{warp2}
M_4 (\Lambda_4=0) & : & \qquad H(z)=z/l_5 \\
\label{warp3}
AdS_4 (\Lambda_4 < 0) & : & \qquad H(z)= l_4/l_5 \sin z/l_4 ,
\eeqa
where $\Lambda_5 =-6/l^2_5$ and the four-dimensional cosmological
constant $\Lambda_4 = \pm 3/l^2_4$ is arbitrary. The metric 
$\gamma_{\mn}$ describes four-dimensional de Sitter, flat Minkowski, and
anti-de Sitter spacetimes, respectively, depending on the warping
factors. These metrics actually describe the same five-dimensional 
anti-de Sitter spacetime with radius $l_5$, and simply correspond to 
different ways of slicing it. If we introduce a 3-brane with uniform 
tension $\sigma$ at $z=0$ perpendicular to the fifth direction, 
one needs cutting and gluing of parts of the $AdS_5$ spacetime
in order to make the geometry smooth around the 3-brane. The resulting 
geometries are still described by the metrics above with replacing
$z \rightarrow |z| +c$. Here $c$ is an arbitrary integral constant which
is related to the location of a 3-brane. This $c$ and the tension of a
3-brane determine the cosmological constant $\Lambda_4$; 
$\left|\sigma\right| =\sigma_0 \cosh c/l_4$ for $dS$ brane,
$\left|\sigma\right|=\sigma_0=3\sr{-\Lambda_5/6}/8\pi G_5$ for flat
brane, and $\left|\sigma\right|=\sigma_0 \cos c/l_4$ for $AdS$
brane~\cite{KR,CBW}. 

Now the Ricci-flat metric embedding in the flat brane world can be 
generalized as follows: The metric given in Eq.~(\ref{metric0})
satisfies the five-dimensional Einstein equation with a negative
cosmological constant $\Lambda_5$ if the metric $\gamma_{\mn}$ is any
solution of the four-dimensional Einstein equation with the cosmological
constant $\Lambda_4$. In particular, one can easily construct 
$AdS$/$dS$ black string solutions by taking the 
$AdS_4$/$dS_4$-Schwarzschild black holes for $\gamma_{\mn}$ such as 
\beq
ds^2 = H^{-2}(z) \Big[ -f(r)dt^2 +\frac{1}{f(r)}dr^2 
+r^2d\Omega_2^2 +dz^2 \Big] ,
\label{bsmetric}
\eeq
where 
\beq
f(r)= 1-\frac{r_0}{r}-\fr{\Lambda_4}{3}r^2 .
\eeq
We wish to study the classical stability of black string solutions
constructed in Eq.~(\ref{bsmetric}) with the warping factors $H(z)$ in 
Eqs.~(\ref{warp1}-\ref{warp3}) under linearized metric perturbations. 
Especially, we will focus on the case of $AdS$ black strings because 
their nature of stability is quite different from those of other two
cases and usual black strings studied in 
Refs.~\cite{GL,GLcharge,Gcosmo,Reall}. Before going into the linearized
analysis in detail, let us give several naive arguments revealing the
basic nature of stabilities in considerations. 

The five-dimensional Riemann tensor squared of black strings in
Eq.~(\ref{bsmetric}) is given by
\beq
R_{MNPQ}(g)R^{MNPQ}(g) \simeq  \fr{10}{9} \Lambda^2_5 
+H^4(z) \Big[ R_{\mu\nu\alpha\beta}(\gamma)R^{\mu\nu\alpha\beta}(\gamma)
-\fr{8}{3}\Lambda_4^2 \Big] ,
\eeq
where $R_{\mu\nu\alpha\beta}(\gamma)$ is the four-dimensional Riemann
tensor constructed from the metric $\gamma_{\mn}$. 
Thus, for the cases of $dS$-Schwarzschild black strings (\ref{warp1})
and Schwarzschild black strings (\ref{warp2}), there generically exist 
curvature singularities at $z= \infty$ in addition to the usual 
singularity at the center of black strings ({\it e.g.}, $r=0$). 
In fact, these are naked singularities, not surrounded by some event 
horizon~\cite{CHR}. For Schwarzschild black strings in particular, 
Gregory~\cite{Gcosmo} has shown that they are unstable, presumably 
indicating fragmentation into an array of black holes. If these black 
strings were stable and so no tendency to fragmentation, the full 5D 
spacetimes would be pathological due to such naked singularities. 
For $AdS$ black strings (\ref{warp3}), however, the function $H(z)$ is 
finite everywhere and so there is no naked or curvature singularity
other than the usual ones at $r=0$. Therefore, we expect that $dS$ black 
strings would be unstable at least near the ``Rindler horizon'' 
$z= \infty$, but $AdS$ black strings would not necessarily have to be so. 

A common and more convincing argument is based on entropy comparison
between different black hole configurations. The existence of black 
string instability is often explained by arguing that there exists 
a length for a segment of black string above which a compact black hole 
with the same mass becomes entropically favourable~\cite{GL,GLcharge}. 
It possibly indicates that a black string decays into an array of black 
holes. We will compare the entropy contained in a segment of the black 
string with that contained in a five-dimensional compact black hole of 
the same mass. Since there is no known exact five-dimensional black hole
solution in the presence of a 3-brane, we consider only the case with no
3-brane. However, we believe the result obtained equally applies to the
case with a 3-brane because the presence of a 3-brane does not change
the important nature of the Kaluza-Klein mass spectrum relevant in the
linearized perturbation analysis as shall be shown below. 

The entropy contained within a segment of $AdS$/$dS$ black string can be
obtained by integrating the area of the horizon as follows,
\beq
S_{\rm b.s.} = \frac{A}{4}=\frac{1}{4}
  \int_{a}^{b}{\cal A}(z)\frac{dz}{H(z)}
  =\pi r_+^2 L, \hs{3ex} L=\int_{a}^{b}\frac{dz}{H^3(z)} .
\label{Sbs}
\eeq 
Here ${\cal A}(z) = 4\pi r^2_+(z)=4\pi r^2_+/H^2(z)$ with $f(r_+)=0$
is the area of the $AdS_4$/$dS_4$-Schwarzschild black hole measured 
by an observer at $z={\rm const.}$. Notice that we can set $l_4 =l_5=l$
by using diffeomorphism. The mass contained in the segment
can also be obtained by integrating the first law of black hole 
thermodynamics $\delta M=T\delta S$ as in Ref.~\cite{EHM}
\beq
 M = \int_{0}^{r_+}T\frac{\del S}{\del r_+}dr_+
  =\frac{r_0L}{2} .
\eeq
Now the {\it five}-dimensional $AdS$-Schwarzschild black hole is
described by
\beq
ds^2 = -f(R)dt^2 +\frac{1}{f(R)}dR^2 +R^2d\Omega_3^2, \hs{3ex}
 f(R)= 1+\frac{R^2}{l^2} -\frac{R_0^2}{R^2} . 
\eeq
Thus, black hole entropy and mass become 
\beq
S_{\rm b.h.} = \frac{\pi}{3}R_+^3, \qquad\qquad  
M = \frac{R_0^2}{4} , 
\eeq
respectively. By identifying the mass with that of the string segment,
one can express $\Delta S =S_{\rm b.s.}-S_{\rm b.h.}$ as a function of 
$r_+$, $l$, and $L$. Surprisingly, this difference can be positive, 
independent of the ``length'' of black string segment $L$ provided that
\begin{eqnarray}
\mbox{} & & dS (\Lambda_4 > 0): \qquad  \frac{\sqrt{13}-3}{2} l 
\simeq 0.30 l < r_+ , \\
& & AdS (\Lambda_4 < 0): \qquad  \frac{3-\sqrt{5}}{2} l \simeq 0.38 l 
< r_+ < \frac{3+\sqrt{5}}{2} l \simeq 2.62 l.
\label{en}
\end{eqnarray}
Note that, for the case of $dS$ black string, the event horizon 
should be inside of its cosmological horizon, $r_+ \leq l/\sqrt{3}
\simeq 0.58 l$ ({\it i.e.}, $r_0\leq 2l/3\sqrt{3} \simeq 0.38 l$).

Thus, it appears that black strings are entropically more favourable 
than the five-dimensional $AdS$-Schwarzschild black hole with the same 
mass no matter how long the hypercylindrical horizon of black string is
extended, possibly indicating stability, if the size of four-dimensional 
horizon $r_+$ lies on the range shown above. For black strings with the
horizon radius $r_+$ not belonging to this range, the black hole becomes
entropically favourable as the ``length'' of string segment $L$
increases as usual. Interestingly, however, the black string segment
again becomes entropically favourable if its ``length'' increases
further. Here, however, we would like to point out that this sort of 
``global'' thermodynamic stability argument should not be taken
seriously since this argument in the viewpoint of the classical black
hole area theorem shows only some plausibility for the classical decay 
of black strings. 

The Gubser-Mitra~\cite{GM} conjecture can be regarded as a refinement of
the ``global'' entropy argument given above, and is proved in 
Ref.~\cite{Reall}. Although black string solutions we consider do not
have a translational symmetry due to the warping factors in
Eqs.~(\ref{warp1}-\ref{warp3}), it will be interesting to apply this
conjecture to our case. The local thermodynamic stability of a segment
of $AdS$/$dS$ black string will be determined by the sign of the heat
capacity given by 
\beq
\fr{dM}{dT} = -2\pi \fr{1-\Lambda_4r^2_+}{1+\Lambda_4r^2_+} r^2_+L .
\label{GM}
\eeq
For the $AdS$ case, one can easily see that the heat capacity is
negative for $r_+ < 1/\sr{-\Lambda_4}= l_4/\sr{3}$, but becomes positive
for $r_+ > l_4/\sr{3}$~\cite{HP}. Thus, we expect $AdS$ black strings
become stable classically when $r_+ > l_4/\sr{3}$ according to the GM
conjecture. On the other hand, $dS$ black strings are expected to be
unstable classically since they are locally thermodynamically unstable
for $r_+ < l_4/\sr{3}$ and the cosmological horizon locates at 
$r_+ = l_4/\sr{3}$.

\section{Linearized perturbation analysis}
\label{linear}

So far, we have given three naive arguments which possibly indicate that
$AdS$ black strings are stable when the four-dimensional horizon radius
becomes large. Now let us perform the classical stability analysis
explicitly. We consider small metric perturbations about $AdS$/$dS$
black string background spacetimes and see whether or not there exists
any mode which is regular spatially, but grows exponentially in time. 
By choosing the Randall-Sundrum gauge~\cite{Gcosmo,RS,GT}, {\it vacuum} 
metric perturbations can be written as follows
\begin{eqnarray}
 ds^2 = H^{-2}(z)
  \left[ (\gamma_{\mu\nu}+h_{\mu\nu})dx^{\mu}dx^{\nu} +dz^2\right] ~,
\qquad  \cov^{\mu}h_{\mu\nu} = 0,
\qquad h=\gamma^{\mu\nu}h_{\mu\nu}=0 ,
\label{gf}
\end{eqnarray}
where $\cov$ is the covariant derivative operator compatible with the
four-dimensional $AdS$/$dS$- Schwarzschild black hole metric
$\gamma_{\mu\nu}(x)$ in Eq.~(\ref{bsmetric}). 
Then the linearised Einstein equations for vacuum metric fluctuations 
become simply
\begin{eqnarray}
  \Box h_{\mu\nu}(x,z) +2R_{\mu\rho\nu\tau}h^{\rho\tau}(x,z)
  -\left(-\del_z^2+\frac{3\del_zH}{H}\del_z\right)h_{\mu\nu}(x,z) =0 ,
\end{eqnarray}
where $\Box \equiv\gamma^{\rho\tau}\cov_{\rho}\cov_{\tau}$ and
$h^{\mu\nu}\equiv \gamma^{\mu\rho}\gamma^{\nu\tau}h_{\rho\tau}$.
Putting $h_{\mu\nu}(x,z) = H^{3/2}(z)\xi (z) h_{\mu\nu}(x)$, this 
equation can be decomposed into the four dimensional part 
({\it e.g.}, massive Lichnerowicz equation) and the fifth part as usual
\begin{eqnarray}
\mbox{} & & \Delta_L h_{\mu\nu}(x)
\equiv\Box h_{\mu\nu}(x) + 2R_{\mu\rho\nu\tau}h^{\rho\tau}(x)
  =m^2 h_{\mu\nu}(x),  
\label{lich}  \\
& & \big[ -\partial_z^2 +V(z) \big]\xi (z) = m^2\xi (z), \hs{3ex}
V(z)=-\frac{3}{2}\fr{H''}{H}+\fr{15}{4}\Big(\fr{H'}{H}\Big)^2 .
\label{KKmass}
\end{eqnarray}

Thus, one sees that essentially the fifth dimension gives massive 
gravitons as usual in the Kaluza-Klein (KK) point of view and 
their mass spectrum can be read off from the form of effective 
potential $V(z)$. For the flat case in Eq.~(\ref{warp2}), $V(z)$ 
vanishes as $z \rightarrow \infty$ and so the KK mass spectrum is 
continuous starting at $m=0$. For the $dS$ case in Eq.~(\ref{warp1}), 
$V(z)$ goes to a non-zero constant, $9/4l_4^2$, and the KK mass 
spectrum is again continuous, but has a non-zero minimum mass, 
$m_{\rm min.} =3/2l_4$. For the $AdS$ case, however, $V(z)$ grows 
infinitely, making effectively a confining box due to $AdS$ nature
of the spacetime. Thus, the $z$ direction is effectively compactified
even if it is still infinite in proper length. Consequently, the KK mass
spectrum becomes discrete now and its lowest mass is $m_{\rm
min.}=4/l_4$. Here one can observe that the scale of effective
compactification is $l_4$ instead of $l_5$. Note that this effective
compactification does not happen when the $AdS_5$ spacetime is sliced 
into four-dimensional Minkowski or $dS_4$ submanifolds. When a 3-brane is 
introduced, it gives a delta-function-like potential well at the 
position of the 3-brane, producing an attractive force. 
Hence the KK mass spectrum for black string in brane world scenarios 
has basically the same feature as that in the case of no 3-brane, 
but its magnitude is somewhat reduced~\cite{KR,Miemiec}. Here we should
point out that the non-zero finiteness of the lowest KK mass for
$AdS$/$dS$ cases plays an important role for the stability of black
strings as shall be shown below explicitly.\footnote{Actually, there
exist massless KK modes as well for both $AdS$ and $dS$ cases. They
are not normalizable except for the $dS$ case with a 3-brane. However,
these massless modes are irrelevant for searching instability modes as
will be explained in more detail below.} 

As explained above, we wish to find any instability mode which is a
solution of the massive Lichnerowicz equation in Eq.~(\ref{lich}) 
with suitable reference to gauge and boundary conditions at the future
event horizon and the spatial infinity. Since higher angular momentum
fluctuation modes are more stable in general, we will consider a zero
angular momentum mode only, an $s$-wave mode~\cite{GL,GLcharge}. 
General spherically symmetric perturbations which give instability can 
be written in canonical form as~\cite{VRW,GL}
\begin{eqnarray}
h_{\mu\nu} (x) &=& e^{\Omega t}
  \left(\begin{array}{cccc}
   H_{tt}(r) &H_{tr}(r) &0 &0 \\
	 H_{tr}(r)&H_{rr}(r)&0&0\\
	 0&0&K(r)&0\\
	 0&0&0&K(r) \sin^2\theta
	\end{array}\right) ~,
\end{eqnarray}
with $\Omega >0$. Using transverse traceless (TTF) gauge conditions 
in Eq.~(\ref{gf}), we can eliminate all but one variable, 
say $H_{tr}$, from the Lichnerowicz equation, obtaining a second order
ordinary differential equation as follows~\cite{GLcharge,full}:
\begin{eqnarray}
\mbox{} & & \left[ \Omega^2+m^2f+\frac{ff''}{2}-\frac{f'^2}{4}
      +\frac{ff'}{r}\right] H''_{tr} 
      \nn\\
& & +\left[ \frac{\Omega^2}{f}\left(3f'+\frac{2f}{r}\right)
     +m^2\left(2f'+\frac{2f}{r}\right)
     +\left(\frac{3f'f''}{2}+\frac{ff''}{r}+\frac{3f'^2}{2r}
       -\frac{3f'^3}{4f}+\frac{2ff'}{r^2}\right)\right] H'_{tr} 
  \nn\\
& & +\left[ -\left(\frac{\Omega^2}{f}+m^2\right)^2
     +\frac{\Omega^2}{f}\left(-\frac{2f}{r^2}+\frac{f'}{r}+\frac{f''}{2}
			 +\frac{5f'^2}{4f}\right)
     +m^2\left(-\frac{2f}{r^2}+\frac{2f'}{r}+\frac{f''}{2}
	  -\frac{f'^2}{4f}\right)
  \right.
 \nn\\
& & \quad\quad \left.
  +\left(\frac{5f'^2}{2r^2}-\frac{f'^3}{rf}+\frac{f'^2f''}{4f}
    -\frac{ff''}{r^2}+\frac{3f'f''}{r}+\frac{f''^2}{2}
    -\frac{2ff'}{r^3}-\frac{f'^4}{4f^2}\right)
  \right] H_{tr} =0 .
\label{mastereq}
\end{eqnarray}

Asymptotically $H_{tr}$ have the following solutions;
\beq
 H_{tr} \sim \left\{\begin{array}{ll} 
                    r^{-5/2 \pm \sqrt{9/4+(ml_4)^2}} & 
\qquad \mbox{for $r \rightarrow \infty$,}  \\
                   (r-r_+)^{-1\pm \Omega /2\kappa}  &
\qquad \mbox{for $r\rightarrow r_+$,}
                    \end{array}
             \right. 
\label{BC}
\eeq
for the $AdS$ case, and
\beq
 H_{tr} \sim  \left\{\begin{array}{ll}
                     (r_{++}-r)^{-1\pm \Omega /2\kappa_{++}} &
\qquad \mbox{for $r\rightarrow r_{++}$,}   \\
                     (r-r_+)^{-1\pm \Omega /2\kappa }  &
\qquad \mbox{for $r\rightarrow r_+$,} 
                     \end{array}
              \right.
\eeq
for the $dS$ case. Here $\kappa = f'(r_+)/2$ and $\kappa_{++} = 
-f'(r_{++})/2$ are surface gravities of the event and cosmological 
horizons, respectively, and $r_{++}$ in the $dS$ case denotes the 
cosmological horizon. As emphasized in Refs.~\cite{VRW,GL,GLcharge}, 
it is very important to impose right boundary conditions on the 
perturbations. Since our analysis is based on linearized equation, 
any fluctuations should remain ``small.'' This seemingly excludes 
asymptotic solutions with negative roots near the horizon. 
However, one can see that even asymptotic solutions with positive root
diverge near the horizon when $\Omega /2\kappa < 1$. It is probably due
to that the Schwarzschild coordinates are not good near the horizon. In
fact, it turns out that, if we use some regular coordinate system such
as Kruskal coordinates, the only asymptotic solution suitable for our
linearized analysis is the one with positive root with any $\Omega$
($>0$) as pointed out in Refs.~\cite{GL,GLcharge}. At the spatial
infinity for the $AdS$ case, if we require vanishing boundary condition
as usual~\cite{HH}, even the positive root satisfies this condition
provided $ml_4 < 2$. Actually, asymptotic $AdS$ spacetimes are not
globally hyperbolic. Thus, one needs to impose some extra condition by
hand in order to make the dynamics of metric fluctuations well
posed~\cite{Wald}. As imposed usually for matter fields in the pure
$AdS$ background spacetime~\cite{AdSbc}, we require the total energy of
gravitational fluctuations on this $AdS$ background should be
conserved. This requirement is satisfied only if $\lambda < -5/2$ when
$H_{tr} \sim r^{\lambda}$ at $r \sim \infty$. Hence only the negative
root is satisfactory in Eq.~(\ref{BC}). We will give the details
in~\cite{full}.

With these boundary conditions described above, we now search
instability modes characterized by $(\Omega , m)$ which are solutions 
of Eq.~(\ref{mastereq}) for given $r_0$ and $l_4$. In other words, 
for $AdS$ case we start from a solution with 
$H_{tr} \sim r^{-5/2 -\sqrt{9/4+(ml_4)^2}}$ at $r \sim \infty$, and find 
$\Omega$ which makes $H_{tr}$ extrapolate to 
$H_{tr} \sim (r-r_+)^{-1 +\Omega /2\kappa}$ near the horizon through 
Eq.~(\ref{mastereq}). Similarly, we start from 
$H_{tr} \sim (r_{++}-r)^{-1 +\Omega /2\kappa_{++}}$ near the cosmological
horizon for the $dS$ case. Since the equation Eq.~(\ref{mastereq}) is 
quite complicate to handle analytically, we solve it numerically. 
It, however, is worth to observe some scaling symmetries in
Eq.~(\ref{mastereq}) as follows:
\begin{eqnarray}
 r \rightarrow \alpha r, \hs{3ex}
 r_0 \rightarrow \alpha r_0, \hs{3ex}
 l_4 \rightarrow \alpha l_4 \,\, (\Lambda_4 
\rightarrow \alpha^{-2} \Lambda_4 ), \hs{3ex}
 \Omega \rightarrow \alpha^{-1} \Omega, \hs{3ex}
 m \rightarrow \alpha^{-1} m .
\end{eqnarray}
So we can fix one of these parameters, say $l_4=1$.  Moreover when 
$r_0 \gg l_4$, $f \sim -r_0/r \pm r^2/l_4^2$, and so there exists
another approximate scaling symmetry for large black holes given
by~\cite{HH}
\begin{eqnarray}
 r \rightarrow \alpha r, \hs{3ex}
 r_0 \rightarrow \alpha^3 r_0, \hs{3ex}
 l_4 \rightarrow l_4, \hs{3ex}
 \Omega \rightarrow \alpha \Omega, \hs{3ex}
 m \rightarrow m.
\end{eqnarray}
Then we find $\Omega \sim r_0^{1/3}$ for the case of large black holes.

\begin{figure}[tbp]
 \centerline{\epsfxsize=85mm\epsffile{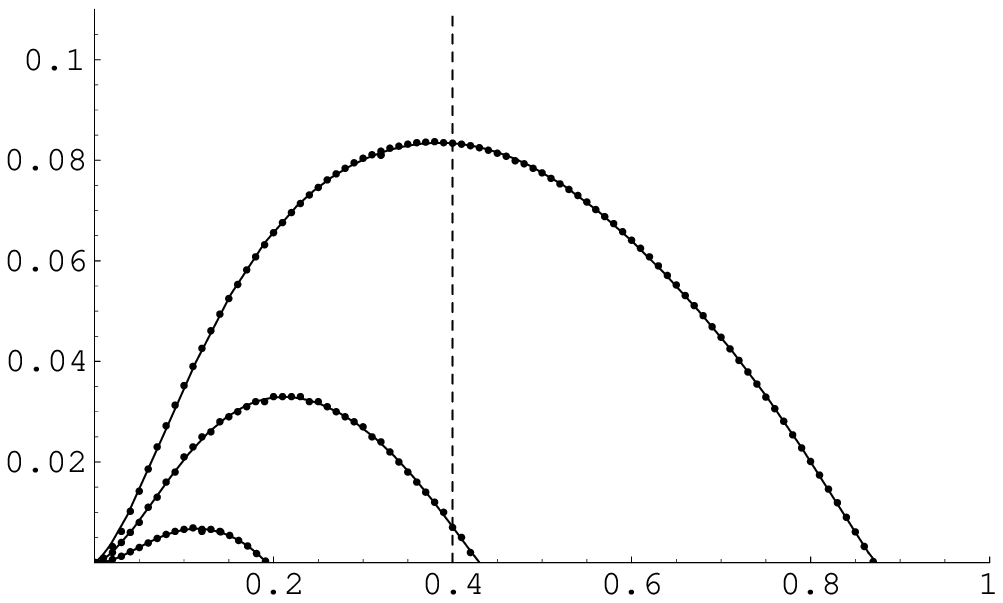}
  \epsfxsize=85mm\epsffile{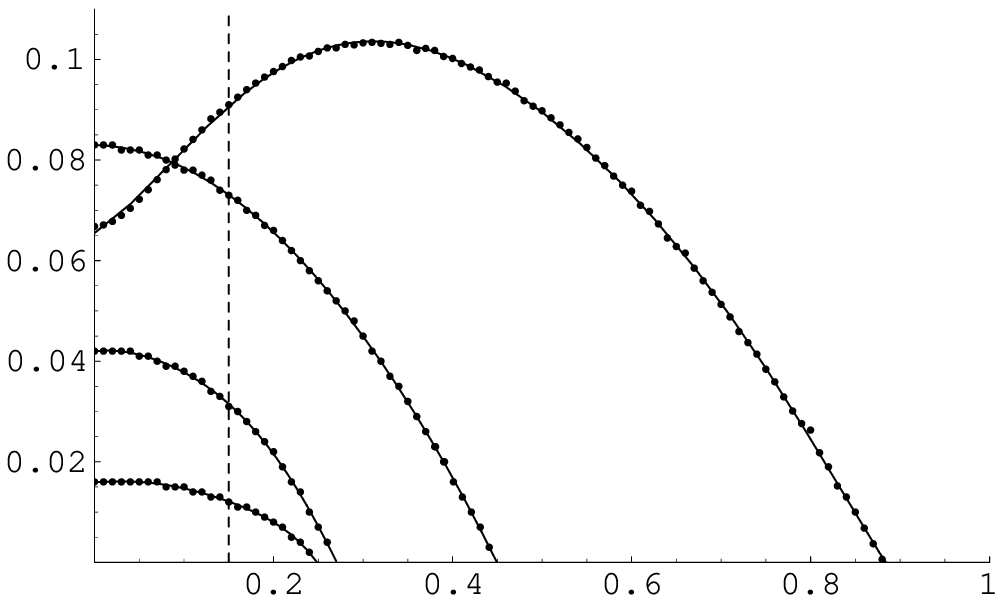}
   \put(-215,135){$\Omega$}
   \put(-130,146){$dS$}
   \put(-15,15){$m$}
   \put(-70,80){$r_0=1$}   
   \put(-140,60){$r_0=2$}   
   \put(-210,64){$r_0=3.5$}   
   \put(-218,34){$r_0=3.8$}   
   \put(-400,146){$AdS$}
   \put(-460,135){$\Omega$}
   \put(-320,78){$r_0=1$}   
   \put(-420,56){$r_0=2$}   
   \put(-440,26){$r_0=4$}   
   \put(-260,15){$m$}
 }
 \caption{The left figure is for the $AdS$ case with $r_0=1, 2$ and $4$.
 The right figure is for the $dS$ case with $r_0=1, 2, 3.5$, and $3.8$. 
 The Nariai solution corresponds to $r_0 \simeq 3.85$. The fixed $AdS$ 
 and $dS$ radius is $l_4=10$. The straight vertical lines denote the 
 lowest KK masses, $0.4$ for $AdS$ and $0.15$ for $dS$.}
 \label{fig1}
\end{figure}

Using Mathematica and Gear method for solving differential equations,
we obtained the results in Fig.~\ref{fig1}. For both $AdS$/$dS$ cases, 
one can see that the instability shrinks in parameter space as the mass 
parameter $r_0$ increases with fixed $AdS$/$dS$ radius. More precisely, 
the mass of the so-called threshold unstable mode~\cite{Reall}, 
$(\Omega , m)=(0, m_*)$, decreases fast down to zero as the horizon 
radius $r_+$ increases for the $AdS$ case whereas it approaches to some 
non-zero finite value as $r_+$ increases towards the cosmological
horizon for the $dS$ case. Thus, it appears that there always exist 
instability modes. However, the KK mass $m$ cannot be arbitrary, but is 
determined by the geometry in the fifth direction through 
Eq.~(\ref{KKmass}) as explained before. The lowest KK masses are $4/l_4$ 
and $3/2l_4$ for the $AdS$ and $dS$ cases without 3-brane, respectively, 
as denoted in Fig.~\ref{fig1} by straight vertical lines. So, if the 
threshold mass becomes smaller than this lowest KK mass for a
certain $r_0$, there exists indeed no unstable mode. As can be seen in 
Fig.~\ref{fig1}, this happens actually when $0.21 l_4 \lesssim r_0$ 
({\it i.e.}, $0.20l_4 \lesssim r_+$) for the $AdS$ case. Therefore, 
we find that the $AdS$ black string is unstable when its
four-dimensional horizon size is small, but it becomes stable
when the horizon size is larger than the order of the $AdS_4$ radius
({\it i.e.}, $r^{\rm cr}_+ \simeq 0.20 l_4$). 
On the other hand, the presence of a 3-brane reduces the lowest KK
mass. Consequently, it increases the value of critical horizon radius
for stable black strings in $AdS$ brane world model. In particular,
in the vicinity of the flat brane world ({\it i.e.}, $\Lambda_4 \sim 0$ 
or $l_4 \sim \infty$), $AdS$ black strings become almost always unstable 
since $m_{\rm min.}\simeq 0$, which can be expected from the results in
Ref.~\cite{KR}.

For the $dS$ case, on the other hand, although the threshold 
mass decreases as the horizon radius increases up to the cosmological
one, they all still seem to remain larger than the lowest KK mass (see 
also Fig.~\ref{fig3}). Therefore, $dS$ black strings seem to be always 
unstable. In particular, the instability seems to persists all the way 
down to the Nariai solution in which the event horizon coincides with 
the cosmological horizon. It, however, should be pointed out that the 
Nariai limit must be treated separately since boundary conditions become
invalid and the numerical error in our analysis increases near such 
extremal case. As argued in Ref.~\cite{GS}, the stability behavior of 
this case might be very different from that of non-extremal cases. The
presence of a 3-brane in $dS$ black strings again makes the system more
unstable since it reduces the lowest KK mass in the unit of $l_4$. 
For the flat case ({\it i.e.}, $\Lambda_4 =0$), we have confirmed the 
results obtained in Ref.~\cite{Gcosmo}. That is, since the KK mass 
spectrum is continuous with zero lowest mass and the threshold mass 
asymptotes to zero as $r_0 \rightarrow \infty$ as can be seen in
Fig.~\ref{fig3}, all black strings are unstable in this case. 

\begin{figure}[tbp]
 \centerline{\epsfxsize=85mm\epsffile{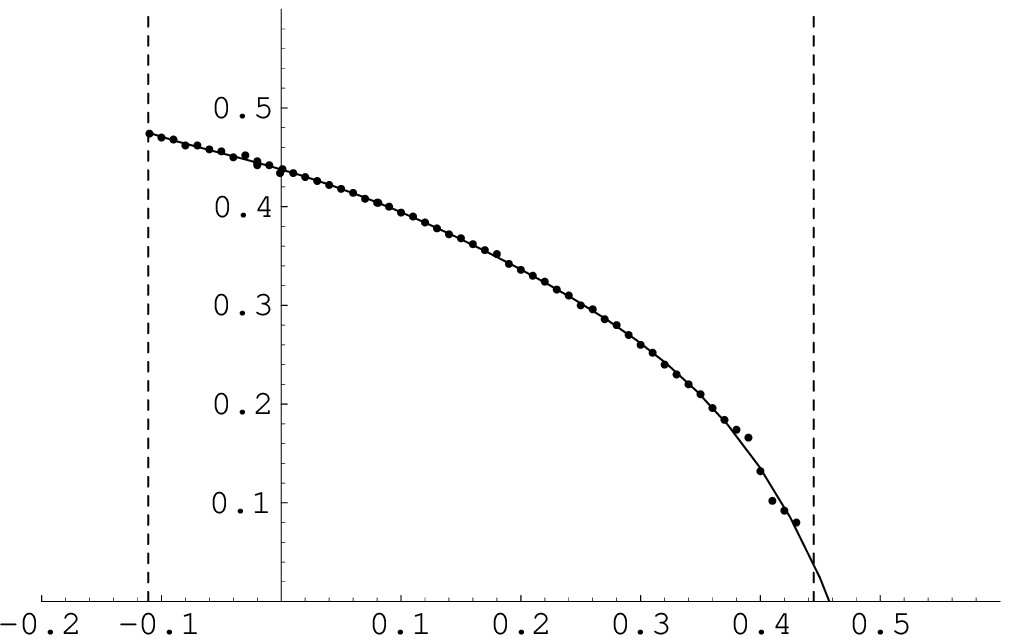}
   \put(-180,155){$m_{*}$}
   \put(-130,20){$AdS$}
   \put(-178,1){$0$}
   \put(-230,20){$dS$}
   \put(5,8){$-\Lambda_4$}
  }
 \caption{The threshold masses $m_*$ for varying $\Lambda_4$ with given 
 $r_0=2$. The left vertical dotted line denotes the Nariai limit and the 
 right one the critical $\Lambda_4$ predicted by the GM conjecture.}
\label{fig2}
\end{figure}

\begin{figure}[tbp]
 \centerline{\epsfxsize=85mm\epsffile{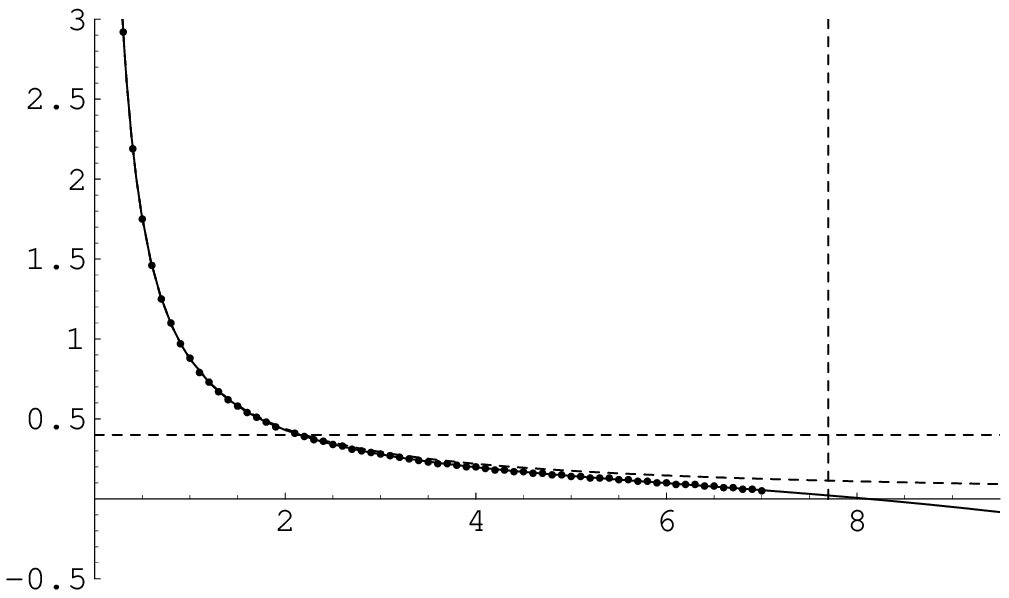}
  \epsfxsize=85mm\epsffile{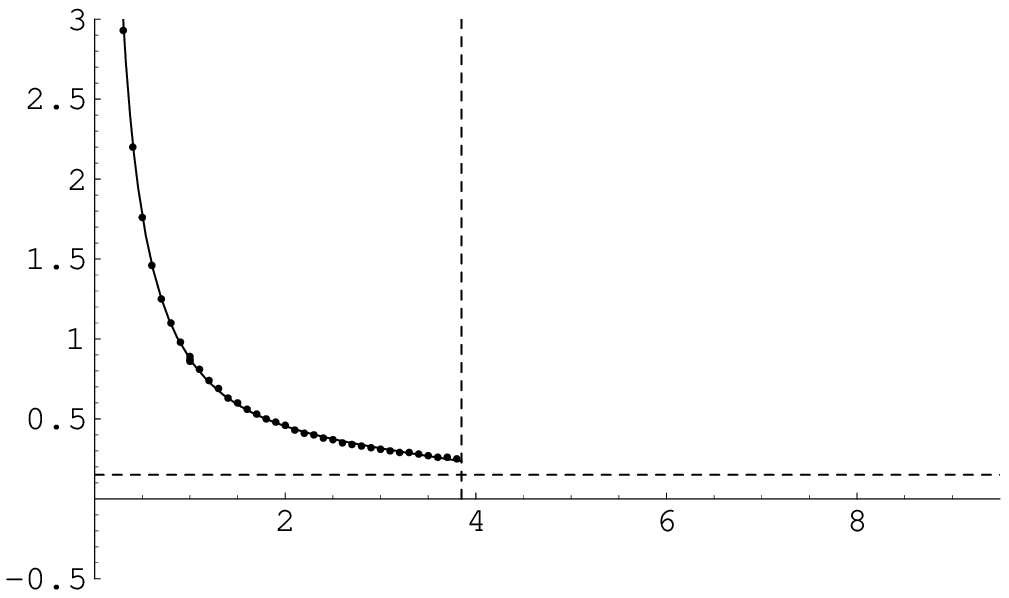}
   \put(-205,135){$m_{*}$}
   \put(-130,146){$dS$}
   \put(-15,25){$r_0$}
   \put(-80,33){\small{$m_{min.}$}}
   \put(-350,43){\small{$m_{min.}$}}
   \put(-400,146){$AdS$/{\rm Flat}}
   \put(-450,135){$m_{*}$}
   \put(-260,25){$r_0$}
 }
 \caption{The left figure: threshold masses for varying $r_0$ with given 
 $l_4=10$ in the $AdS$ case. The vertical dotted line denotes the critical 
 $r_0 \simeq 7.7$ predicted from the Gubser-Mitra conjecture. 
 The numerical data stops at $r_0 \simeq 7.0$. The right figure: same 
 diagram for the $dS$ case. The vertical dotted line denotes the Nariai 
 limit, $r_0 \simeq 3.85$. Note that $m_*(r_0\simeq 3.85)=0.29 > 
 m_{\rm min.} \simeq 0.15$.}
\label{fig3}
\end{figure}

Fig.~\ref{fig2} illustrates how the threshold mass changes for a 
given $r_0$ as the cosmological constant $\Lambda_4$ varies away 
from zero. It shows that the instability in parameter space shrinks 
as $\Lambda_4$ becomes negative ({\it i.e.}, $AdS$ case), but expands as 
$\Lambda_4$ becomes positive ({\it i.e.}, $dS$ case). In other words, 
basically adding negative cosmological constant has a stabilising 
effect as in the case of adding charge into black 
strings~\cite{GLcharge} whereas adding positive cosmological constant 
gives destabilising influence.

It will be interesting to see how well the results obtained by 
explicit perturbation analysis agree with those in naive arguments 
given before. For the $AdS$ case, critical values for stable
black strings were predicted as $r_+ \simeq 0.38 l_4, 0.58 l_4$ in
Eqs.~(\ref{en}) and (\ref{GM}) by the entropy comparison and by the 
GM conjecture, respectively. The numerical results predict 
$r_+ \simeq 0.20 l_4$ which agrees within the order of one. 
The entropy comparison, however, also predicts another critical horizon
radius, $r_+ \simeq 2.62l_4$, across which black strings become unstable
again. We have searched various parameters around this critical value,
but could not find any unstable black string. Thus, our numerical
results agree well, at least qualitatively, with the GM conjecture, but
with the entropy argument only in part. For the $dS$ case, on the other 
hand, only the prediction in GM conjecture agrees well with the 
numerical results. 

Fig.~\ref{fig3} shows how the threshold mass for a given $\Lambda_4$
decreases as the black hole becomes large. These numerical results agree
qualitatively well with those in Refs.~\cite{thermo} obtained
analytically with some approximation and different gauge choices in a
different context. Both flat ({\it i.e.}, $\Lambda_4=0$) and $AdS$ ({\it
i.e.}, $\Lambda_4 < 0$) cases give almost same decreasing pattern for
small $r_0$, but they start to deviate as the black hole becomes large,
around $r_0 \simeq 4$. As $r_0 \rightarrow \infty$ for the flat case,
$m_*$ denoted by the dotted curved line in Fig.~\ref{fig3} asymptotes to
zero ($\sim 1/r_0$), but never touches it. Consequently, since the
continuum KK mass spectrum starts at $m=0$, one can see again that all
black strings are unstable no matter how large $r_0$ is. If the black
string is compactified, however, the continuum KK mass spectrum becomes
discrete. The massless mode is not a real instability mode, but
presumably a gauge artifact~\cite{GLcharge,GL} since the Lichnerowicz
equation with $m=0$ becomes that of pure four-dimensional black
holes. So, the lowest instability mode will start at non-zero $m$. Then
Fig.~\ref{fig3} shows that compactified black strings in the flat case
will become stable if $r_0$ is larger than some critical value
determined by the compactification scale. The stability of $AdS$ black
strings can be understood similarly in this point of view. It is because
the $AdS_5$ nature of geometry in the fifth direction with $AdS_4$
slicing gives an effective compactification whose scale is determined by
$l_4$ instead of $l_5$ as explained above. However, we point out
there is another interesting feature in this case. As can be seen in
Fig.~\ref{fig3} for the $AdS$ case, although numerical error increases
as $m_*$ becomes small, the curve for $m_*$ seems to touch the
horizontal axis if the data points are extrapolated further. Moreover,
this terminating point seems to agree with the critical value $r_0
\simeq 0.77l_4$ ({\it i.e.}, $r_+ \simeq 0.58l_4$) obtained by the GM 
conjecture, the horizon radius across which the heat capacity changes 
its sign. Consequently, one can expect that the black string will be 
stable at least if $r_0$ is larger than this terminating value, no
matter what the KK mass spectrum is. Therefore, in addition to the 
stabilization due to an effective compactification, $AdS$ black string 
solutions seem to have a sort of intrinsic tendency for stabilization 
probably due to the $AdS_4$ nature of longitudinal four-dimensional 
geometries. 

It can be noticed that the critical value, $r_+ \simeq 0.20l_4$, 
obtained for $AdS$ black strings does not exactly agree with, but occurs  
a bit ``earlier'' than that of the GM conjecture, $r_+ \simeq 0.58l_4$. 
This discrepancy, however, is expected because the GM conjecture assumes 
non-compact translational symmetry. Actually, this condition can be
easily replaced in the proof of the GM conjecture~\cite{Reall} as
follows: a black string/brane, as long as its KK mass spectrum is
continuous starting at zero mass, is classically stable if, and only if,
it is locally thermodynamically stable. One can see that the flat case
satisfies this modified GM conjecture. The GM conjecture then predicts
that a terminating point must exist if the system is locally
thermodynamically stable. In fact, our numerical results show  not only
that the terminating point exists, but also that it agrees with the
critical value of local thermodynamic stability in the GM
conjecture. Thus, it can be expected that a black string having discrete
KK mass spectrum becomes stable, if it happens, before the terminating
point as in the $AdS$ case. 

Finally, we have so far concentrated on the feature of stability with
special emphasis on stable black string configurations in $AdS_5$
spacetimes. Now it will be also interesting to see what the final states
would be for unstable black strings. In order to answer this question,
we just need to know how the eigenfunction $\xi (z)$ in
Eq.~(\ref{KKmass}) with given $m$  behaves along the fifth coordinate
$z$. For black strings in the flat case, it has been argued in
Ref.~\cite{Gcosmo} that the interval of successive wiggles in proper
length becomes exponentially tiny towards the $AdS_5$ horizon, and so
the string is somewhat stable near the 3-brane but quickly becomes
unstable away from it, generating an accumulation of ``mini'' black
holes towards the $AdS_5$ horizon. For $dS$ case with 3-brane, the shape
of the potential $V(z)$ in Eq.~(\ref{KKmass}) is a volcano type and
similar to that of the flat case. The only difference is that $V(z)$
approaches to a non-zero constant as $z \rightarrow \infty$ ({\it e.g.},
the ``Rindler'' horizon) instead of vanishing. Then $\xi(z)$ will be
similar to that of the flat case which is Bessel function, but goes to
zero more quickly. Accordingly, the feature of fragmentations will be
almost same as that in flat case, with a bit stronger instability. 
For the $AdS$ case, however, it turns out to be very different. 
The potential $V(z)$ is again a volcano type around the 3-brane, 
but diverges at the boundary of $AdS_5$ ({\it e.g.}, the conformal
infinity), effectively creating a box. Thus, $\xi(z)$ will behave
like the Hermite function which is an eigenfunction of a harmonic
oscillator with slight modifications in the vicinity of the 3-brane.
Consequently, the black string becomes again stable near the boundary of
$AdS_5$ as well as in the vicinity of the 3-brane, generating multi
black holes in between. This is why a segment of $AdS$ black string
becomes entropically favourable again when its length $L$ in
Eq.~(\ref{Sbs}) becomes large enough. 

\section{Conclusion}

To conclude, we have shown that, although black strings in $AdS$
spacetimes which are not locally asymptotically flat are generically
unstable classically under linearized metric fluctuations, the $AdS$
black string solutions are stable when the longitudinal size of the
horizon is larger than the order of $AdS_4$ radius. Generically, 
adding negative cosmological constant has a stabilization effect whereas
adding positive cosmological constant has a destabilization influence. 
It will be straightforward to extend our study to higher dimensional
cases. We believe the essential feature of stability for $AdS$ black 
string/brane solutions in higher dimensions will be same.

\section*{Acknowledgments}

Authors would like to thank H. Kodama and T. Tanaka for helpful
discussions. GK also would like to thank R. Gregory, N. Ishibashi, 
T. Jacobson, M. Natsuume and R.M. Wald for useful discussions 
and communications. TH thanks T. Goto and K. Ishikawa for helpful 
advice concerning Mathematica. This work was supported by JSPS 
(Japanese Society for Promotion of Sciences) Postdoctoral 
Fellowships.

\newcommand{\J}[4]{#1 {\bf #2} #3 (#4)}
\newcommand{\andJ}[3]{{\bf #1} (#2) #3}
\newcommand{\AP}{Ann.\ Phys.\ (N.Y.)}
\newcommand{\MPL}{Mod.\ Phys.\ Lett.}
\newcommand{\NP}{Nucl.\ Phys.}
\newcommand{\PL}{Phys.\ Lett.}
\newcommand{\PR}{Phys.\ Rev.\ D}
\newcommand{\PRL}{Phys.\ Rev.\ Lett.}
\newcommand{\PTP}{Prog.\ Theor.\ Phys.}
\newcommand{\hep}[1]{ hep-th/{#1}}
\newcommand{\hepp}[1]{ hep-ph/{#1}}
\newcommand{\hepg}[1]{ gr-qc/{#1}}

\end{document}